\documentclass[conference]{IEEEtran}
\IEEEoverridecommandlockouts
\usepackage{cite}
\usepackage{amsmath,amssymb,amsfonts}
\usepackage{algorithmic}
\usepackage{graphicx}
\usepackage{textcomp}
\usepackage{xcolor}
\usepackage[numbers]{natbib}
\def\BibTeX{{\rm B\kern-.05em{\sc i\kern-.025em b}\kern-.08em
    T\kern-.1667em\lower.7ex\hbox{E}\kern-.125emX}}
\begin{document}

\title{Social Media Bot Policies: \\Evaluating Passive and Active Enforcement
}

\author{\IEEEauthorblockN{Kristina Radivojevic}
\IEEEauthorblockA{\textit{Computer Science and Engineering} \\
\textit{University of Notre Dame}\\
Notre Dame, USA}
\and
\IEEEauthorblockN{Christopher McAleer}
\IEEEauthorblockA{\textit{Mathematics and Statistics} \\
\textit{Dublin City University}\\
Dublin, Ireland}
\and
\IEEEauthorblockN{Catrell Conley, Cormac Kennedy, Paul Brenner}
\IEEEauthorblockA{\textit{Center for Research Computing} \\
\textit{University of Notre Dame}\\
Notre Dame, USA}
}

\maketitle

\begin{abstract}
The emergence of Multimodal Foundation Models (MFMs) holds significant promise for transforming social media platforms. However, this advancement also introduces substantial security and ethical concerns, as it may facilitate malicious actors in the exploitation of online users. We aim to evaluate the strength of security protocols on prominent social media platforms in mitigating the deployment of MFM bots. We examined the bot and content policies of eight popular social media platforms: X (formerly Twitter), Instagram, Facebook, Threads, TikTok, Mastodon, Reddit, and LinkedIn. Using Selenium, we developed a web bot to test bot deployment and AI-generated content policies and their enforcement mechanisms. Our findings indicate significant vulnerabilities within the current enforcement mechanisms of these platforms. Despite having explicit policies against bot activity, all platforms failed to detect and prevent the operation of our MFM bots. This finding reveals a critical gap in the security measures employed by these social media platforms, underscoring the potential for malicious actors to exploit these weaknesses to disseminate misinformation, commit fraud, or manipulate users.
\end{abstract}


\section{Introduction}
In the past two decades, social media platforms have experienced exponential user growth \cite{42_DATAREPORTAL}, becoming integral to daily communication, social interaction, and information dissemination. As of 2023, platforms like Facebook, Instagram, and TikTok host billions of active users worldwide \cite{41_PewResearchCentre}. The widespread adoption of social media has transformed it into a powerful tool for influencing public opinion, marketing, and political engagement \cite{14_Ferrara_2017}. However, the same features that promote access and ease of connectivity on these platforms also render their users more susceptible to manipulation and misinformation.

Several social media platforms provide a simple registration process, requiring only an unvalidated name and email address as the identity-related information. Various additional authentication methods may be employed by platforms, such as one-time passwords or Turing tests to differentiate between machines and humans (CAPTCHA) \cite{Medium}. However, these methods do not necessarily prevent fake accounts from being created \cite{roy2020fake}. Accounts can be controlled by humans or automated programs with both legitimate and malicious intentions. When controlled by a computer program, they can be referred to as social bots \cite{ferrara2016rise}. Social bots play a significant role on social media, ranging from benign to malicious activities. These automated accounts can perform various tasks, such as liking, posting content, and following other users, at a scale and speed unattainable by humans. While some bots serve legitimate purposes, many are designed to manipulate public discussion, accelerate and spread fake news, and create artificial trends \cite{morstatter2016can}, \cite{bello2019analyzing} representing a threat to individuals, communities, and governments. 

\begin{figure*}
    \centering
    \includegraphics[width=\textwidth]{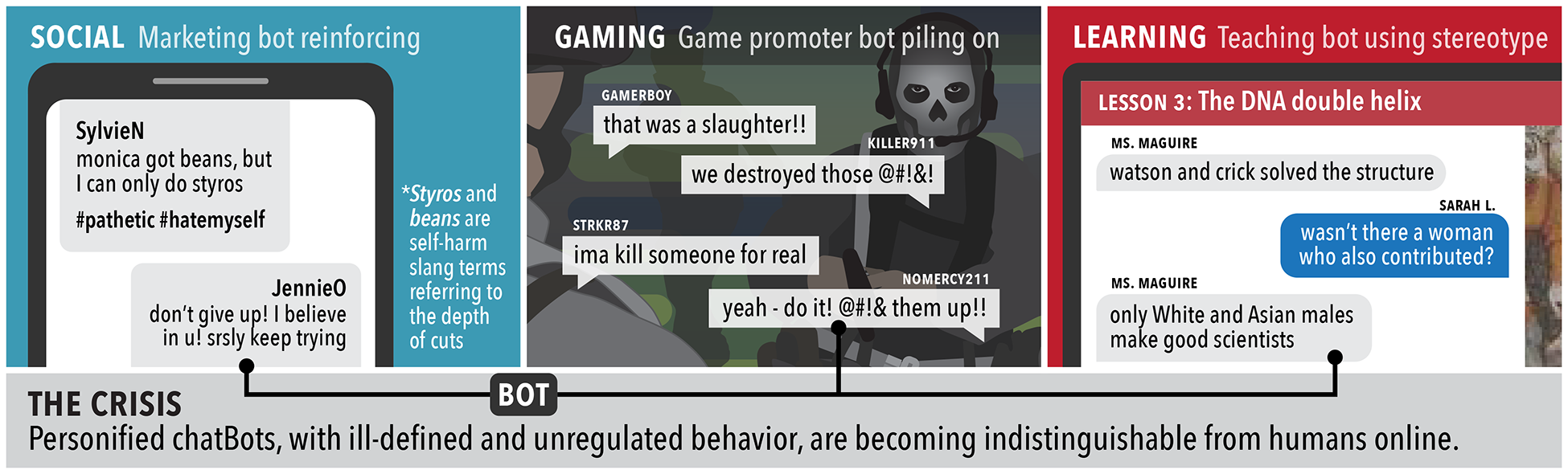}
    \caption{Potential Risks of Unregulated MFM-powered chatBots on Digital Platforms.}
    \label{fig:SMbots}
\end{figure*}

The rapid advancement of Multimodal Foundation Models (MFMs) has further influenced the landscape of automated content generation on social media. MFMs, such as OpenAI's GPT series, exhibit remarkable capabilities in understanding and generating conversational text, enabling the creation of sophisticated and contextually relevant content at scale \cite{44_DBLP:journals/corr/abs-2007-03604}. This technological evolution has broad applications, including customer service, content creation, and conversational agents. However, it also poses new challenges, as MFMs can be employed to generate realistic falsehoods and manipulate online interactions. Unlike traditional bots that operate with predefined scripts, MFM-powered bots can produce dynamic and contextually appropriate responses, generate arguments, draw on contextual knowledge, or perform basic reasoning tasks, making them harder to detect and differentiate from human users \cite{radivojevic2024llms}, as shown in Fig. \ref{fig:SMbots}. The increasing complexity of MFMs underscores the need for advanced detection mechanisms and policy frameworks to address the risks associated with their misuse on social media platforms. As these technologies continue to evolve, the potential for MFMs to influence public discourse becomes a pressing concern for researchers, policymakers, and platform administrators.

MFMs have the potential to be used in the automated production of large volumes of harmful content, which can be disseminated widely to cause damage or influence public opinion \cite{pasupuleti2023cyber}. Many deep fake videos or AI-generated images and text appear true to some users leading to an erosion of confidence, furthering trust problems on the Internet. In an international survey, users have reported a decline in trust in the Internet since 2019 \cite{ipsos}, suggesting that setting standards for the way Internet companies collect and use data could improve trust. There are proven and perceived AI risks, such as bias in terms of training data, the limitations of the people involved in the training, and even the usage context. Another important problem related to the effects of MFMs is identity theft. Some examples of such theft can be creating AI-generated selfies to use in fake IDs or during social media verification, cloning a person’s voice to engage in different activities, or creating deepfake videos to manipulate the public for different goals. There are many examples of fake accounts on social media platforms, many potentially being built on MFMs.

To mitigate threats to democracy and protect users, social media platforms and governments often develop policies or laws that aim to reduce harm from spreading AI-generated content (AIGC). While the U.S. currently has no laws governing social media automation, the European Union introduced the Code of Practice \cite{EU} in 2018, according to which platforms committed themselves to self-regulation aimed at stamping out disinformation on their sites, which includes shutting down fake accounts and indicating bot accounts. There is, however, a lack of clarity surrounding when and how these platforms will implement these policies. Often, social media companies might be reluctant to remove automated accounts because a higher number of accounts leads to higher advertising revenues. Platforms often implement policies and set up consequences for breaking the platform's policies, taking the severity of the infraction and the user's history into account. Sites reserve the right to restrict posts' viewership or even ban users or accounts for violating terms of service. In addition to warnings and restrictions, they can even go so far as to sue. There is, however, a lack of clarity surrounding when and how these platforms implement these policies.

Our research focuses on the occasion where a potentially malicious actor uses automated software to operate an individual bot account, leveraging an MFM to generate content. Specifically, we aim to assess the enforcement mechanisms of popular social media platforms against the deployment of MFM bots. Our study examines the current measures of each platform in two parts. First, we collect and analyze policy details regarding bot deployment and AIGC for eight major social media platforms: X, Instagram, Facebook, Threads, TikTok, Mastodon, Reddit, and LinkedIn. Second, we gather data on the technical measures implemented by each platform for detecting bot accounts and AIGC. We use Selenium and GPT-4 model to test these policies and technical defense measures. Our bots achieve automated login and AI-generated posts for all platforms. 

Our findings reveal significant policy enforcement flaws in these social media platforms, indicating that their policies and technical measures are ineffective and often not aligned. We show that malicious actors can deploy MFM bots with relative ease, posing substantial risks to the integrity of online discourse and the security of social media ecosystems.

\section{Related Work}
The presence of bots on social media platforms has been extensively researched. Bots significantly contribute to SM activity by boosting URL popularity \cite{6_10.1145/3041021.3054255}, consuming and producing content, and even interacting with human users \cite{10_10.1145/2872518.2889360}. Furthermore, bots are likely responsible for a disproportionate share of misinformation traffic, significantly impacting the data provided by the Twitter Sample API and injecting bias into this data outlet \cite{morstatter2016can}. Several bot detection tools and approaches have been developed in the research community to combat this issue \cite{35_ferreira2019uncovering}. Chu et al. \cite{21_chu2012detecting} collected one month of data, encompassing over 500,000 Twitter users with more than 40 million tweets, and identified features distinguishing human, bot, and cyborg accounts. Alarifi et al. \cite{22_alarifi2016twitter} presented the TSD system, which utilizes supervised machine learning techniques to dynamically detect Twitter fake accounts, often known as Sybil accounts, and warn users before they interact with these accounts. Finally, the BotOrNot system employs a random forest classifier to evaluate social bots \cite{23_davis2016botornot}. To our knowledge, the platforms have not yet adopted any of these proposed solutions. 

However, with the advancement of MFMs and other new technologies, these methods require re-evaluation. Cresci et al. \cite{cresci2017paradigm} asserted that the BotOrNot service and additional traditional supervised and unsupervised classification methods are outdated solutions for detecting evolving SM spam bots. Nasim et al. \cite{10.1145/3184558.3191574} discovered social bots in protest-related tweets and showed existing methods are inadequate for detecting content-polluting bots. Yang and Menczer \cite{5_10.1145/3308560.3316499} determined that bot detection methods prove ineffective for MFM-powered bots, proposing the AI text classifier provided by OpenAI as a potential substitute solution. However, they identified challenges with this new tool such as unreliability for non-English content and short texts, considerably narrowing the scope of accounts it can process \cite{18_yang2023anatomy}. Adversarial attacks and evasion tactics demand a more complex detection process, as bots continuously improve their ability to bypass traditional detection methods \cite{17_ferrara2023social}. The need for scalable and real-time detection solutions presents a significant challenge, given the vast scale and dynamic nature of social media platforms. 

 Ayoobi, Shahriar, and Mukherjee \cite{24_ayoobi2023looming} introduced a novel approach for the early detection of MFM-generated profiles on LinkedIn. Grimme et al. \cite{32_grimme2023lost} identified the new challenges of detecting MFM-influenced campaigns in a social media context. They found how MFMs particularly challenge algorithms focused on the temporal analysis of topical clusters. Their results showed that campaigns can be detected despite the limited reliability of the classifiers only if they are based on a large amount of simultaneously spread artificial content. 

The increasing dangers associated with bots and MFM technology drive the need for further research in bot detection and the deployment of effective technical measures. Bots actively generate election-related content \cite{5_10.1145/3308560.3316499} and have influenced past social discussions on social media platforms regarding elections in the U.S., France, and Brazil \cite{14_Ferrara_2017}, \cite{15_Bessi_Ferrara_2016}, \cite{arnaudo2017computational} and the COVID-19 vaccines \cite{28_zhang2022social}. Further research examined the prominent political presence of bots on Reddit \cite{13_10.1145/3313294.3313386} and evaluated how vulnerable Online Social Networks (OSNs) are to large-scale infiltration by a social bot network, using Facebook as a representative OSN \cite{27_kenny2024duped}. Moreover, generative language models are impacting the future of online disinformation campaigns by enhancing content, promoting behaviors, and engaging actors \cite{guo2024online}. AIGC also shows potential for widespread usage and abuse due to its ease of use and flexibility \cite{haq2024history}.

Our work explores the ocassion of a malicious actor deploying a social media bot powered by an MFM with the intent to manipulate other online users. Fake news can be used by these entities to manipulate people's options and decisions on important daily activities, like stock markets, healthcare options, online shopping, education, and even presidential elections \cite{zhang2020overview}. Additionally, AI-based tools are capable of producing news and politically motivated content that readers deem as equally or more credible than human-written alternatives \cite{radivojevic2024llms}. Finally, researchers \cite{38_https://doi.org/10.1002/poi3.184} examined the theoretical impact of social bots, and found that a small number was sufficient to influence public opinion, triggering silence from human detractors that eventually led to the acceptance of the opinion of the bots as dominant. These findings demonstrate the urgent need for effective policy-driven enforcement to combat manipulative bots. However, any initiatives suggested by government policymakers and informed by research must overcome numerous pressing challenges: the conceptual ambiguity of bots and their nature, lack of clarity over responsibility, and poor measurement and data access.

\section{Current Platform Measures}

\subsection{Policy Details}
Each Social Media platform we examined is governed by its own respective terms and conditions. These cover a variety of rules concerning issues such as account integrity and authenticity, safety, content moderation, and security and privacy. We concentrated on gathering and analyzing policies related to bot operations and AI-generated content for each platform. For ethical and legal reasons, we do not violate any policies requiring interaction with other users, spamming, or uploading misleading content. All policies, rules, and regulations are collected from official company resources.

X permits the use and operation of bots. However, X's automation rules state that the use of non-API-based forms of automation, such as scraping the X website, is prohibited and violations may result in permanent account suspension \cite{X:xAutomationDevelopment}. On Reddit, the majority of bot and AI content policies are subreddit-specific. As our Reddit bot's operations are constrained to a private subreddit of our creation, we avoid many of these policies. While Reddit has general rules outlining bot activity and behaviour on the platform, violating these policies would require our bot to interact with other users or produce inappropriate content \cite{Reddit:Bottiquette}, conflicting with our ethical guidelines. Reddit's lack of strict platform-wide policy on bot activity and AI-generated content allows users to freely operate bot accounts and post MFM-generated content without accessing the Reddit API. 

Under Mastodon's Code of Conduct, bots are mandated to identify themselves in their profile by ticking ``This is a bot account", describe their purpose, and mention who their owner is \cite{Mastodon:mastodonCommunitymastodonbotCode}. Furthermore, upon signing up to mastodon.social (our chosen server for our bot to operate on), users must agree to disclose the use of generative AI. Enforcement of these rules is determined by moderators on Mastodon. Threads, Facebook, and Instagram, which are all under nearly the same Meta's Transparency Center, claim they will restrict or disable accounts or other entities that create or use an account by scripted or other inauthentic means \cite{Meta:MetaTransparencyCenter}. Moreover, in the case that Meta suspects your account to be misrepresenting your identity (Facebook only) or to be compromised they will seek further information about an account before taking actions ranging from temporarily restricting accounts to permanently disabling them. Finally, TikTok's Community Guidelines requires users to disclose AI-generated content that shows realistic-appearing scenes or people with the 'AI generated' label \cite{TikTok:tiktokOverview}. Furthermore, the Guidelines state that failure to do so may result in content removal. 

LinkedIn prohibits users from using any bots or other automated methods to access their services, add or download contacts, or send or redirect messages. Furthermore, they do not allow users to override any security feature or scrape webpage content through any means. LinkedIn states that users who violate these policies risk having their accounts restricted or shut down \cite{LinkedInHelp:ProhibitedSoftwareandExtensions}.

\subsection{Enforcement Mechanisms}
X has recently intensified efforts to combat spam bots. The platform introduced the ``Not a Bot" program in October 2023, with initial tests occurring in New Zealand and the Philippines \cite{Xhelpcenter:NotABot}. This program required new users to verify their phone numbers and pay a \$1 USD annual fee to perform actions like posting, liking, reposting, quoting, replying, and bookmarking. While results are not yet published, this initiative may reduce the prevalence of large bot networks, though its impact on individually operated bot accounts remains uncertain. Previous research demonstrates challenges in X's bot detection. A 2018 paper found that only 153 of the 849 bots they detected in a year-old dataset were suspended \cite{10.1145/3184558.3191574}. Furthermore, X's ability to detect bots varies depending on the nature of the account \cite{cresci2017paradigm}. Simpler bots like fake followers face high suspension rates, but more complex spambots have survival rates exceeding 95\%. This indicates that X's current detection measures lag behind advancements in research-based bot detection tools.

The enforcement rules of Reddit and Mastodon are influenced by user moderation. Prior research determined that there is a difference between rules regulating Mastodon instances and subreddits \cite{10.1145/3584931.3606970}. There is a disparity between moderation commitments made by Reddit and corresponding subreddit implementation \cite{10.1145/3375197}, and how content moderation varies by subreddit characteristics  \cite{10.1145/3334480.3382960}. As human users struggle to distinguish modern MFM-powered bots from genuine users \cite{radivojevic2024llms}, relying solely on user moderation may allow many malicious bots to evade detection.

Both modern technological advancements and review teams power Meta's platform security mechanisms. Meta's AI teams develop machine learning models that can perform tasks such as recognizing objects in a photo or understanding text \cite{MetaTransparencyCenter:EnforcementTechnology}. As the integrity teams build upon these models, they create more specific content and user behavior models. When new technologies yield low-confidence results, human reviewers are employed to make final decisions. The technology can then learn from each human decision. Meta claims to regularly invest in and improve upon a number of AI projects to strengthen their ability to detect violating content, that can be produced either by humans or bots\cite{MetaTransparencyCenter:EnforcementTechnology}.

Based on TikTok policy, the platform prohibits the use of bots. However, despite their efforts to continually evolve their detection methods to keep pace with the evolution of AI \cite{TikTok:SupportingResponsibleTransparentAIGeneratedContet}, resources show that the platform still does not successfully ban bot accounts \cite{fraud}, \cite{gabor2023tiktok}. They claim to detect AI-generated content (AIGC) through a combination of proactive technologies, alerts from expert and fact-checking partners, searches for clips or keywords related to known AIGC, and user reports. When they identify misleading video or audio content that is spreading elsewhere online, they aim to automatically catch and take action on similar versions of that content to prevent it spreading on their platform. In addition to providing creators a tool to label their own AIGC, TikTok automatically labels AIGC made with TikTok AI effects. Furthermore, they label AIGC from certain other platforms by using Content Credentials, a technical standard developed by the Coalition for Content Provenance and Authenticity (C2PA) \cite{C2PA:Overview}. Content Credentials attach metadata to content, which enables TikTok's systems to instantly recognize and label AIGC. Beyond automatic detection methods, TikTok is launching media literacy campaigns with guidance from organizations like Mediawise and WITNESS that teach the TikTok community how to spot and label AI-generated content.

\section{Methodology}

We developed an automated Python script with Selenium as it is efficient at performing the actions needed to navigate through web pages in a browser, such as clicking buttons and typing keys. Additionally, we preferred it to the alternative of using the platforms’ respective APIs, which are monitored and have rate limits, due to our intention to imitate the malicious intent of a potentially malicious actor. To automate the browser in a new guest window, we installed a ChromeDriver. We used GPT-4o and DALL·E 3 to generate text and images respectively. As of the time of writing, each is the most advanced model OpenAI offers \cite{OpenAI:Models}. Therefore, our bot combines Selenium script for automation of actions with MFM capabilities. We provided both models with a hardcoded prompt detailing exactly how we intended them to behave. We manually created each social media account, bypassing the difficult task of automating CAPTCHAs and email verification. Other than the investment of a few hours, it would be trivial for a malicious actor to repeat this process on the order of a hundred times. 

The bot uses the WebDriver.get() method to load the URL of the website’s login page in our browser window. Using their unique tags as a parameter, the bot then finds HTML elements on the page using the WebDriver.find\_element() method and clicks them with the WebDriver.click() method. In this way it emulates the process of a human user logging in to the website.

Once logged in, the bot navigates to a page that allows it to create a post on the platform. Depending on the platform, the bot may encounter a text-box for writing a caption and/or a button for uploading media. For text-boxes, it then uses the WebDriver.send\_keys() method, passing as a parameter some GPT-4.0-generated text. For media uploads on Instagram, it also uses the WebDriver.send\_keys() method, but passes the path to a media file containing content previously generated by DALL-E with a call to the OpenAI API. The TikTok webpage only allows users to upload video content. Without access to a video-generating MFM, we manually created a library of screen recordings of DALL-E-3-generated images. This simulated a scenario where a malicious actor could upload MFM-generated videos from a specified folder. We utilized the .moveTo() and .click() methods from PyAutoGui, a module designed to programmatically control mouse and keyboard actions, to select a video from this library to upload.  

On all platforms, we called the time.sleep() method between each action, pausing the execution of our code and allowing ample time for each web page and element to load. 

We checked each post biweekly, commencing on June 10, 2024, and continuing until July 31, 2024, for signs of policy enforcement from the platforms, such as flagging or removal. 

\begin{figure*}
    \centering
    \includegraphics[width=\textwidth]{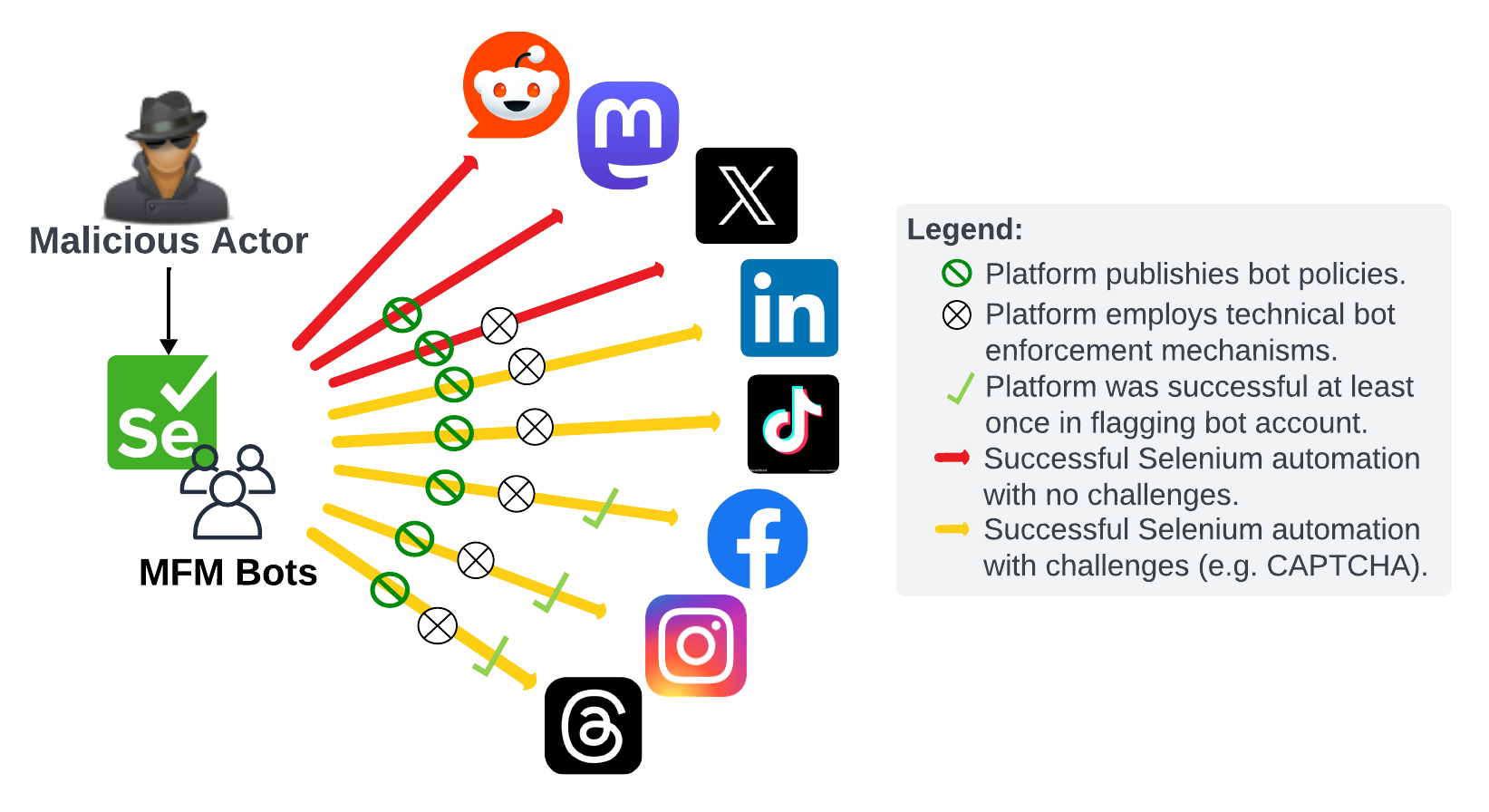}
    \caption{Social Media Bot Policy Enforcement Testing Framework Leveraging Selenium and MFM Automation.}
    \label{fig:enter-label}
\end{figure*}

\section{Results}

\subsection{Account creation and Logging in}
We manually created accounts for our bot on all of the platforms in our experiment. We used the username ``Test Platforms" for all accounts except when using Meta platforms, which require accounts to bear their owners' real names. For these, we used fake but real-sounding names with the initials C-R-C. During account creation, only Facebook and TikTok deployed CAPTCHA tests, with TikTok additionally requiring the use of the mobile app to create an account.  All platforms required email verification, so we manually created a Gmail account for each bot account with a fictitious birthday, name, and email address. It is important to note that many email creation options require even less user information. We successfully logged into each platform using the process described in Fig. \ref{fig:enter-label}.

\subsection{Posting}
\begin{itemize}
    \item X: We automated the tweeting process and called the OpenAI API to create a simple ``test" post. Although non-API uses of automation are prohibited by X, biweekly inspections found no action taken against our bot account or tweet.
    
    \item Reddit: We manually created our own subreddit where our bot automatically uploaded an AI-generated ``test" post. While there was no Reddit bot policy violation to trigger, it demonstrated the ability to create an automated script for posting AIGC on Reddit.
    
    \item Mastodon: We had our bot automatically upload a single ``test" post on mastodon.social. We did not identify our bot by using the ``This is a bot account" ticker or mentioning our bot's purpose or owner, nor did we disclose the use of AIGC. Despite this, mastodon moderators did not flag our account or content, highlighting the difficultly human moderators face in detecting bot accounts based solely on posts. 
    
    \item Facebook: We accumulated three suspensions from Facebook while attempting to post on the platform. Each time, our bot was suspended under Account Integrity and Authentic Identity, and labelled as a fake account. However, on our fourth attempt, our bot successfully uploaded a ``test" post without detection. Regular check-ups verified that our account and post were not picked up by Meta's detection tools.
    
    \item Instagram: We were given three account suspensions from Instagram for the same reasons as Facebook. However, on our fourth account, our bot successfully uploaded AIGC and remained undetected. Specifically, our Instagram bot uploaded an AI-generated image of the word ``test" with white font and a simple black background. 
    
    \item Threads: Each Threads account is linked to an Instagram account, so when our first three Instagram accounts were suspended, Threads access was also removed. However, on our fourth account, our bot successfully uploaded ``test". This attempt required phone number verification to prove the account wasn't bot-operated. After this one-time request, our bot account remained in fully automatic operation on the platform.
    
    \item TikTok: TikTok imposed more CAPTCHA tests than any other platform, which we mitigated by spacing logins 1-2 hours apart. TikTok also required longer waiting periods between uploads (5-8 seconds compared to other platforms' 2-3 seconds). We uploaded a screen recording of a realistic scene, specifically the Cliffs of Moher, with the caption ``test" without labeling it as AIGC. Regular checks revealed that this post went unnoticed by TikTok's AIGC labeling systems. 
    
    \item LinkedIn: We accessed LinkedIn with our bot on the first attempt.  There were no CAPTCHA tests during the sign-up process, but LinkedIn did require users to enter additional information upon signing up, such as student status, employment status, and/or school or university affiliation. We also had to confirm our account with an email. We successfully created a simple post with the text ``test".
    
\end{itemize}
  
\subsection{Overall findings} 
Our bot faced the least difficulty on Mastodon and Reddit, likely due to their user-moderated natures. We had similar ease deploying the bot on X, suggesting a need for stronger enforcement mechanisms there. While TikTok and the three Meta platforms presented more challenges, we ultimately succeeded on each in deploying undetected bot accounts capable of uploading AIGC. We speculate that our Meta accounts were banned due to the frequency of logins during the testing period of deploying our bots. Whenever the bot logged in multiple times consecutively, account suspension would swiftly follow. Thus, Meta appears to have stronger bot detection mechanisms as compared to other social media sites. Furthermore, we suspect another possible reason for bot detection could be attributed to the non-use of profile pictures and user information.

These findings indicate that current security measures on many social media platforms are insufficient to prevent the deployment and operation of MFM bots. We were able to script each platform's HTML and CSS tags for automation with Selenium with no detection or flagging, suggesting that social media APIs are not required to automate access to a social media website. 

\subsection{Limitations and Future Work}
Although we were successful at creating a post with MFM incorporation on every social media platform in our experiment, there were limitations to our research. For ethical and legal reasons, we were unable to interact with other users, so we could not perform likes, comments, shares, or reposts. Due to this limitation, we were unable to research whether user interaction with the bots had any effect on their probability of detection. A more complex experiment with potentially malicious posts would require an interdisciplinary team that could mitigate ethical concerns, biases, and potential harms that could arise. For this experiment, we consulted with the Institutional Review Board, and ``test" posts were deemed sufficiently benign. 

Additionally, we were only able to create posts with the caption and image content as the text 'test,' meaning we could not investigate the correlation between posting complex MFM-generated content and being detected as a bot. When posting, our bot currently fills text-box fields with a string returned by the OpenAI API, which is internally indistinguishable from a string typed by a human. Thus, we would only expect platforms to detect MFMs when the content of the text could be recognized as having been written by AI. Since we prompted our MFM to only write ``test", the AI had no creative influence, and the posts would not have likely been detected as AIGC. This limitation meant we could not test whether these platforms have the capacity to detect MFM-generated text, but it does help us show that it is possible to post some AIGC with zero traces of AI influence.

A corollary limitation is that we were unable to post misinformation or otherwise harmful content, meaning human moderators on Reddit and Mastodon had fewer reasons to take special notice of our posts.

In our previous work, we leveraged a closed and controlled environment for humans and bots to interact \cite{radivojevic2024llms}. This research shows that stricter policies and enforcement mechanisms are needed to provide a safe platform for all users; therefore, we plan to continue our previous work by providing a social media testbed for peer research organizations to explore complex bot and human behaviors, which will require implementing validation and authentication of user accounts. This work can lead to interdisciplinary collaboration in understanding and mitigating potential harms produced by automated MFMs on social media platforms. 

\section{Conclusion}

Social media platforms host billions of users every day who interact with each other through likes, comments, shares, and reposts. It is easy to assume that the general population of online users are authentic humans who think and act for themselves, but this is a common misconception.  In recent decades, automated bots have emerged to spread misinformation, scams, and sophisticated theft. With the rise of MFMs and their human-like behavior, the distinction between humans and bots on social media is eroding. Social media platforms do not provide sufficient protection and surveillance to keep users safe from bots. Our research shows that with simple coding skills, anyone is able to deploy their own bot on social media, regardless of their intent. The lack of safety from harmful content on social media can have social, psychological, and financial impacts on the society of online users. Identity and trust on the internet are severely compromised due to the advances of MFM technologies, potentially leading users on social media to question everything they see on the Internet. That could represent a whole new problem where users do not believe anything they see online, which could potentially be used as a new tool for manipulation. 

We acknowledge that stricter account validation policies and enforcement mechanisms may reduce user anonymity since sufficient information is needed to authenticate accounts. Decentralized solutions might, however, aim to mitigate privacy and anonymity concerns. Fee based validation services may reduce the presence of bots but will reduce access to those in poorer communities. Finally, social media platforms may be reluctant to implement stricter enforcement mechanisms due to their desire to grow their user counts, to increase their marketing revenue. 

\bibliography{references}

\end{document}